# K-12 Teaching and Physics Enrollment

**Samina S.Masood**

Department of Physics

University of Houston Clear Lake

Houston, TX 77058

## Abstract

We have collected and analyzed the relevant data from public schools in greater Houston area of Texas. Based and analyzed. Since the data is only limited to a few school, we are still working on getting more data so that we can compare and contrast the results adequately and understand the core of the enrollment issue at the national level. However, based on the raw data and partial analysis, we propose a few recommendations towards the improvement of science education in Texas Schools, in general, and greater Houston area schools in particular. Our results indicate that the quality of science education can be improved significantly if we focus on the improvement of high school education or even intermediate schools when students are first time exposed to science in a little technical way. Simply organizing teacher training programs at K-12 level as school education plays a pivotal role in the decrease in physics enrollment at the higher level. Similar analysis can actually be generalized to other states to find out the best way to increase the physics enrollment.

## Motivations

In this era of science and technology, it is a big concern that STEM (Science, Technology, Engineering and Mathematics) disciplines are not as popular as they should be in the upcoming generations. Situation in USA is worse than some of the other countries. The urge to be financially independent at the early age, easy-going attitude and the education system may all be blamed for decreasing enrollment in more involved sciences such as physics and chemistry. It is not difficult to notice that a large majority of existing students in STEM disciplines come from Asian communities. Within the STEM disciplines, Physics is badly losing enrollment and attraction. There is a noticeable decrease in physics majors in colleges. Especially undergraduate student enrollment is not good because we do not have too many international students at the undergraduate level. Graduate students population is maintained by international students. We need to investigate the reasons why students lose interest in physics as well as try to look for the ways to attract students in physics. Before getting in to the details we will like to mention some of the key portions of student's comments that we have heard during last decade.

## Why Do We Need to Attract More Students in Physics Program

The biggest issue at this point is that we do not have enough Physics teachers to serve the need of Physics teachers. The nation's schools employ more than 21,000 physics teachers. Two-thirds of them do not have a degree in physics or physics education. The most common degree in science is biology where only two semesters of general physics is required. Most of the science teachers teach physics with biology degree. These students usually cannot do well in physics. Thus there is a need of many more trained physics teachers. Preparing out-of-field teachers is a much faster approach to fulfil the immediate need of physics teachers. Training new teachers will be a part of our long term goals. Senior teachers with physics or physics education degrees may have studied physics many years ago. They may be very good teachers and are perhaps the only ones who are playing a major role in maintaining our physics enrollment. However, these teachers will be better prepared to face their professional challenges in attracting students to physics if they were aware of new developments in the subjects and are exposed to new teaching techniques. After teaching thirty years in colleges at different levels in different countries and talking to a large number of students in physics courses, we heard certain common comments from students such as

- I like physics but I cannot find a good job with Physics.
- Physics is a very difficult subject, I cannot do it.
- I like physics but mathematics is difficult.
- Physics is just mathematics and I don't like mathematics.
- I want to get a quick degree but Physics needs to work hard.
- Physics job market is low. Then why to work so hard and spend so much money and do not get any better job.

These comments clearly indicate that one of the causes of lower enrollment in Physics is the image of the degree itself. Basically, the image of physics as a subject is not spoiled in a day. There are several reasons behind it. These reasons are not limited to but include high school teaching methodology, curriculum designing, testing method and recruitment of teachers in schools. Moreover, it has been noticed that in most of the teacher's certification programs the minimum requirement to get into the program is the same as the minimum requirement of getting in the school . Afterwards it just requires passing grades (C or above) in any subject. This is not attracting our best physics students to a teaching career as a majority of the students with good grades can get into other higher-paying careers. Those who could not go to another career would go into teaching as an alternative. This obviously affects those subjects, which need to be taught by people who are really interested and knowledgeable in the subject itself and can teach it adequately to attract children to physics.

We have looked at the salary structure of schools in Texas [6]. The difference of salary between a bachelor and master degree is just $1,000 dollars per annum and with Ph.D. teachers only get $200 extra in a year as compared to the master degree. In this situation, why would a teacher spend extra time for Ph.D. or even masters? So, there are no real incentives for teachers to go for higher education.

## Recommendations

In this situation we would recommend that to increase physics enrollment at the undergraduate level or to produce more experts in Physics to fulfil our future needs, we have to focus on physics teaching at K-12 level. For this purpose we should

- Provide highly trained science teachers at K-12 level.
- Trained and experts in Physics at high school level.
- High school curriculum should be improved and more conceptual understanding is needed.
- Testing should be heavily based on multiple choice questions. Descriptions and essays should be part of testing procedure.
- Class demonstrations and hands-on experience with lab equipment will help to develop interest
- Semester long projects will be very helpful. Even inter-disciplinary projects for students will create future scientists
- Student's involvement in research through neighboring universities is very helpful to attract students in science disciplines.

Interested readers can consider Ref.[1,2] for further readings.

## References

1. `The Decrease in Physics Enrollment`, **arXiv: physics/0509206**
2. `Better Physics Teaching Can Increase Physics Enrollment', **arXiv: hep-physics/0702089**